\documentclass[aps,prb,amsmath,amssymb,superscriptaddress,showpacs,twocolumn,floatfix]{revtex4} 
\usepackage{amsfonts, relsize, textcomp}
\usepackage{graphics}
\usepackage{graphicx}
\usepackage{hyperref}

\begin{document}

\title{Classification of massive and gapless phases in bilayer graphene}

\author{Bitan Roy}

\affiliation{National High Magnetic Field Laboratory, Florida State
University, Florida 32306, USA}

\date{\today}

\begin{abstract}
I here classify of all the fully gapped \emph{massive} and the \emph{gapless} phases in bilayer graphene. The effective low-energy theory in bilayer graphene is constructed, and various discrete and continuous symmetries of the non interacting system is analyzed. Spinless fermions, placed in a quantizing magnetic field, are considered. The quantum anomalous Hall insulator is properly defined. Constructing a particle-hole doubled 16 component Nambu-Dirac spinor, I recognize all the possible fully gapped and gapless states, which, on the other hand, split the parabolic dispersion into two anisotropic Dirac-like conical ones. A thorough symmetry analysis of all the ordered states is performed. Altogether there are eight insulating and four superconducting phases in bilayer graphene, that can lead to a fully gapped spectrum. Among the gapped superconductors, \emph{three} are spin singlet, which include uniform $s$-wave and two spatially inhomogeneous, translational symmetry-breaking Kekule superconductors. The triplet pairing exhibits an $f$-wave symmetry. Besides the gapped phases, there are eight semimetallic and eight gapless superconducting states in total, available for fermions to condense into. I also find interesting gapless superconducting states, which break the translational symmetry, dubbed as ``gapless-Fulde-Farrell-Larkin-Ovchinikov" superconductors. I also discuss the role of the Coulomb interaction, and propose various experimental tools to determine the nature of the underlying ordered states.      
\end{abstract}


\maketitle

\vspace{10pt}

\section{Introduction}

Carbon based layered materials opened a new frontier in condensed matter physics following the successful fabrication of the single and the bilayer graphene.\cite{geim} The low-energy excitations in monolayer graphene is described by massless, chiral Dirac fermions around the six corners of the first Brillouin zone. However, only \emph{two} out of them are inequivalent.\cite{semenoff}  On the other hand, bilayer graphene with Bernal stacking, leads to parabolic dispersions around those points.\cite{novoselov, castro-neto} Placed in a \emph{weak} quantizing magnetic field both monolayer and bilayer graphene exhibit quantized plateaus of Hall conductivity at integer fillings: $\nu=\pm (4 n + 2)$ in mono-layer and $\nu=\pm (4 n + 4)$ in bilayer graphene, with $n= 0, 1, 2, \cdots$.\cite{IQHE-monobi} The extra four fold degeneracy of all the Landau levels including the zeroth one arises from the valley and the spin degrees of freedom, and the plateaus at $\nu= \pm 2 \;(\pm 4)$ accounts for the fact that only half of the zeroth Landau level is filled in neutral mono(bi)layer graphene. The additional two-fold orbital degeneracy of the zeroth Landau level in bilayer graphene is due to the quadratic dependence of the energy with the momentum.\cite{falko} In the presence of spatially modulated fields, even though the Landau level structure disappears, leading to a continuous spectrum, a finite number of states always persists at zero energy. Density of states at zero energy is twice as much as the magnetic flux enclosed by the bilayer graphene system, while it is simply proportional to the magnetic flux enclosed by monolayer graphene.\cite{Kailasvuori, inhomogeneous}

In a neutral graphene (either bilayer or monolayer), the valence band is completely filled, whereas the conduction band is totally empty, leading to gapless Dirac (in monolayer graphene) or \emph{Dirac-like} (in bilayer graphene) quasi particle excitations in the vicinity of the Dirac points or \emph{valleys}. In pristine monolayer graphene, any weak electron-electron interaction is irrelevant due to the vanishing density of states at the charge-neutrality point, corresponding to a large domain of attraction for the non interacting Gaussian fixed point.\cite{igor-physics} On the other hand, in bilayer graphene the density of states is finite ( in fact, constant). Consequently, even weak electron-electron interactions can be \emph{relevant} in a bilayer graphene.\cite{vafek-kun} Although, in reality such parabolic touching of the bands is achieved at the cost of a fine tuning, where all the hopping amplitudes other than the in-plane nearest-neighbor and out-of-plane direct/dimer ones are neglected. Therefore interactions need to be sizable even in bilayer graphene in order to place the system in any ordered phase. Nevertheless, the requisite strength of the interactions for ordering in bilayer graphene is expected to be sufficiently smaller than that in mono-layer graphene, as the remote hopping amplitudes are weak enough. A similar parabolic band structure can also be realized in checkerboard and Kagome lattices, with particular hopping structures.\cite{kivelson} Otherwise, depending on the relative strength of various components of the finite ranged interaction (repulsive or attractive), fermions in bilayer graphene can condense into a plethora of insulating, gapless (nematic or smectic) and superconducting ground states.

The range and profile of the Coulomb interaction in bilayer graphene can possibly be tuned to certain degree by changing the gate configuration\cite{comment-interaction}, for example. In a recent work\cite{robert}, it has been argued that, interactions are relatively long-ranged in the presence of a single gate, whereas it becomes considerably short-ranged with an additional gate (top and bottom gate). It is, therefore, quite conceivable to observe different ordered phases in bilayer graphene, by changing gate configuration, substrate and thereby tuning the profile and/or range of the Coulomb interaction. The exact dependence of the nature of the interaction on various mesoscopic parameters of the system, however, lacks clear understanding at this moment, and will no longer be part of the present discussion. Nevertheless, it is quite fascinating to note that the bilayer graphene has already been found to support various ordered phases: for example, the \emph{nematic order}, which breaks the parabolic band touching into two anisotropic Dirac like dispersions.\cite{manchester-nematic, yacoby} Besides the gapless nematic state, fully gapped insulating bilayer graphene has also been reported in Refs.~\onlinecite{yacoby, weiss-PRL}, Although the exact nature of the insulating order in these experiments is not quite clear. Furthermore, the $B1$ sample in Ref.~\onlinecite{weiss-PRL} discerns metallic behavior. Rather recently, a \emph{layer anti-ferromagnet} order is shown to persist even down to zero magnetic field. The size of the gap otherwise increases monotonically with the strength of the magnetic field.\cite{Lau} Even more recently\cite{weiss-PRB}, the spin symmetry of the insulating state has been analyzed by studying the evolution of the gap in magnetic field with different orientations, proposing the layer anti-ferromagnet or the \emph{quantum spin Hall insulator} as the viable gapped states in insulating bilayer graphene. Yet another insulating order can be realized by applying an electric field between two layers. It produces a density imbalance of the charge carriers among two layers and that way leads to a gapped spectrum, named as \emph{layer polarized state} \cite{castro-volt-mass}. This phase has also been predicted to arise from electron-electron interactions \cite{Macdonald, levitov-nandkishore-LP}. The insulating gap, reported in Ref.~\onlinecite{Lau}, decreases in the presence of a weak perpendicular electric field of either polarity. This observation ruled out the layer polarized state, as an ordered state in bilayer graphene. In light of the recent experimental observations, it appears that the exact nature of the broken symmetry phases, and their connection with the system's mesoscopic environment remains far from being settled, and raises some natural questions. For example, what are the possible ordered phases (either fully gapped or gapless) available for fermions in bilayer graphene to condense into? How do the ordered phases evolve under the influence of quantizing magnetic fields? What are the spin triplet and/or singlet superconducting state in bilayer graphene? Can any gapless superconducting state be found in bilayer graphene?

In bilayer graphene, the low-energy \emph{Dirac-like} Hamiltonian is eight dimensional, arising from two layers, two valleys and two projections of spin. In monolayer graphene, the Dirac Hamiltonian is also eight dimensional, however, the layer degrees of freedom gets replaced by the isospin or the sublattice ones. If therefore, the Cooper pairs are to be accommodated, the Nambu's particle-hole doubled Hamiltonian is 16 dimensional. I here show that there are altogether $28$, 16-dimensional matrices (${\cal M}$) that anti commute with the low-energy non interacting Hamiltonian, and yield non-zero expectation values of the fermionic bilinears $\Psi^\dagger {\cal M} \Psi$. This number is restricted by the Fermi-Dirac statistics, which here translates to a set of algebraic constraints that the matrices (${\cal M}$) need to satisfy. These orders upon acquiring finite expectation values lead to fully gapped spectrum of the quasi particles. Out of the 28 bilinears, 16 define the insulating orders: layer polarized state (1), layer anti-ferromagnet (3), quantum anomalous Hall insulator (1), quantum spin Hall insulator (3), ``odd" and ``even" kekule currents in the spin singlet $(2 \times 1)$ and triplet $(2 \times 3)$ channels. The characterizations ``odd" and ``even" of the Kekule orders reflect their transformation under the exchange of two Dirac points. Numbers in the parentheses indicate the requisite number of matrices to define the corresponding order  parameter. The remaining $12$ bi linears define three spin singlet and one spin triplet fully gapped superconducting phases. The singlet superconducting orders are conventional $s$-wave and two spatially inhomogeneous \emph{Kekule superconductors}. The latter ones break the translational symmetry of the honeycomb lattice into the Kekule pattern and offer examples of Fulde-Ferrell-Larkin-Ovchinikov (FFLO) type of superconducting phases appropriate to the bilayer graphene. \cite{fulde, larkin} The Kekule superconductors can be either even or odd under the exchange of two valleys. The only fully gapped, triplet superconductor exhibits an $f$-wave symmetry, changing sign six times around the Brillouin zone.

Besides the fully gapped insulating and superconducting orders, several gapless orders, which upon acquiring finite expectation values, split the parabolic dispersion into two anisotropic conical ones, can be realized in bilayer graphene. The separation among these two points is proportional to the size of the gapless orders. \emph{The defining property of any gapless order}, we are interested in here, \emph{is that it anti commutes with only one of the matrices appearing in the non interacting Hamiltonian, while it commutes with the other one}. In conjunction with this definition, a set of algebraic constraints restricts the number of gapless order parameters in bilayer graphene to 56. Thirty two of them take place in the particle-hole channel, while the remaining 24 occur in the particle-particle sector. Altogether, there are eight semimetallic and eight gapless superconducting states. I also show that some gapless superconductors can break the translational symmetry, and I name them as ``gapless-Fulde-Farrel-Larkin-Ovchinikov" superconductors. A subset of all the possible ordered states, which I discuss here, has also been considered recently by Nandkishore and Levitov \cite{nandkishore-classification}.

Organization of rest of the paper is as follows. In the next section I discuss the lattice model of the fermion's hopping and the non interacting spectrum in bilayer graphene. Section III is devoted to arrive at the effective low-energy theory in bilayer graphene, and discuss various discrete and continuous symmetries of the non interacting description. The behavior of various mass orders for spinless fermions in quantizing magnetic fields is discussed in Sec. IV. A 16-component Nambu-Dirac spinor, preserving the spin rotational symmetry is constructed in Sec. V. All the possible insulating and fully gapped superconducting orders are shown in Sec. VI. The classification of all the gapless orders in bilayer graphene is presented in Sec. VII. I discuss the role of the electron-electron interactions in Sec. VIII. Possible experimental probes to determine the nature of the underlying broken symmetry phases in bilayer graphene are proposed in Sec. IX. I summarize the central results and discuss some related issues in Sec. X. The underlying Clifford algebra of the order parameters in bilayer graphene is constructed in the Appendix.

\section{Lattice model of free fermions}

To set the problem up, let us begin our discussion with a simple tight binding description of the free fermions in a bilayer honeycomb lattice. One can decompose it as 
\begin{equation}
H_0=H^{\perp}_0 + H^{\parallel}_0.
\end{equation}        
Here
\begin{equation}
H^{\parallel}_0= t \sum_{j=1,2} \sum_{\vec{A},i} u^\dagger_j (\vec{A}) \;  v_j (\vec{A} -(-1)^j \vec{b}_i) + H.c.,
\end{equation}
corresponds to intra-layer hopping among the sites of the two triangular sublattices. For simplicity, I here suppress the spin degrees of freedom. The intralayer hopping amplitude is $t \sim 2.5$ eV.\cite{gloor}  $u^\dagger_j (\vec{A})$ is the fermion creation operator at one of the triangular sublattices generated by the linear combination of basis vectors $\vec{a}_1=(\sqrt{3},-1)a$ and $\vec{a}_2=(0,1)a$. $v(\vec{B})$ is the fermion annihilation operator on the other sublattice, then located at $\vec{B} = \vec{A}+\vec{b}$, with the vector $\vec{b}$ being either $\vec{b}_1=(1/{\sqrt{3}},1)a/2,\vec{b}_2=(1/{\sqrt{3}},-1)a/2$ or $\vec{b}_3=(-1/{\sqrt{3}},0)a$. $j=1,2$ corresponds to two layers, whereas, the interlayer hopping Hamiltonian takes the following form
\begin{equation}
H^\perp_0 \;=\; H^\perp_{0,0} \;+\; H^\perp_{0,1} \;+\; H^\perp_{0,2},  
\end{equation}
where
\begin{eqnarray}
H^\perp_{0,0}\;&=&\; t_\perp \sum_{\vec{A}} u^\dagger_1 (\vec{A}) \; u_2 (\vec{A}) + H.c., \\
H^\perp_{0,1}\;&=&\; t^{12}_{AB} \sum_{\vec{A},i} u^\dagger_1 (\vec{A}) \;  v_2 (\vec{A} - \vec{b}_i) + H.c., 
\end{eqnarray}
and
\begin{equation}
H^\perp_{0,2}\;=\; t^{12}_{BB} \sum_{\vec{B},i} v^\dagger_1 (\vec{B}) \;  v_2 (\vec{B} - \vec{b}_i) + H.c. .
\end{equation}
Various components of the inter-layer hoppings have been measured experimentally\cite{martin}. Currently, their estimated strengths are $t_\perp \sim 0.3$ eV, $t^{12}_{AB} \sim 0.1$ eV, $t^{12}_{BB} \sim 0.3$ eV. To diagonalize $H_0$, let us define a four-component spinor, 
\begin{equation}
\Psi_k \;=\; \left( u_1 (\vec{k}), v_1 (\vec{k}), u_2 (\vec{k}), v_2 (\vec{k}) \right)^\top.
\end{equation}   
The tight-binding Hamiltonian in this basis is
\begin{equation}
H_0\;=\;
\left( \begin{array}{cccc}
0 & t \; f(k) & t_\perp & t^{12}_{AB} f(k)  \\
t \; f^*(k) & 0 & 0 & t^{12}_{BB} f(k)  \\
t_\perp & 0 & 0 &  t \; f^*(k) \\
t^{12}_{AB} f^*(k) & t^{12}_{BB} f^*(k) & t \; f(k) & 0  \\
\end{array} \right),
\end{equation}
where 
\begin{equation}
f(k)\;=\; \sum_{i=1,2,3} \exp\left({i\; \vec{k} \; \cdot \; \vec{b}_i} \right), 
\end{equation}
and $f^*(\vec{k})$ is the complex conjugate.

Next I drop the remote hopping amplitudes, $t^{12}_{AB}$ and $t^{12}_{BB}$, for the sake of simplicity. Then the particle-hole symmetric energy spectrum is composed of \emph{four} bands with the dispersions
\begin{equation}
E_1 (k) \;=\; \pm \frac{1}{2} \; \; \bigg( \frac{t^2 \; |f(k)|^2}{t_\perp} \bigg),
\end{equation}
and
\begin{equation}
E_2 (k) \;=\; \pm \frac{1}{\sqrt{2}}\sqrt{ 4\; t^2 \; |f(k)|^2\; +\; t^2_\perp \:+\: O\left( |f(k)|^4\right)}.
\end{equation}
Note that $f(k)$ is zero at $\vec{K}=(1,1/\sqrt{3}) 2\pi/a \sqrt{3}$ and $-\vec{K}$, located at the two inequivalent corners of the Brillouin zone. Near the Dirac points, $E_1(k)$ vanishes and the dispersion is comprised of two parabolic bands, touching each other. On the other hand, the spectrum of $E_2 (k)$ is gapped everywhere, and near the Dirac points, the band gap is $\sim 2 \; t_\perp$.\cite{castro-neto} Such parabolic degeneracy can only be achieved after setting the remote hopping terms to zero. The inclusion of the remote hopping $t^{12}_{BB}$ splits the parabolic bands into \emph{four} Dirac cones. This term is also known as ``trigonal warping"\cite{falko, nilsson}. In the rest of the discussion I will not consider the effect of the trigonal warping, unless mentioned. A similar splitting of the parabolic band touchings can also arise due to the Rashba spin-orbit coupling\cite{cristiane}. In general the eigenfunctions of $E_2$ have finite overlap on both the sublattices. However, in the vicinity of the Dirac points they can be considered to be localized on the sites $A_i$, with $i=1,2$. An expansion of $f(k)$ near $\pm \vec{K}$ yields
\begin{equation}
f(\pm \vec{K}+\vec{q})=\frac{3}{2} \; \left( \pm q_x \;+\; i \; q_y\right) + {\cal O} (q^2).
\end{equation}
Therefore
\begin{equation}
E_1 (\pm \vec{K}+\vec{q})\;=\; \frac{1}{2\;m} \; \left( q^2_x \;+\; q^2_y \right),
\end{equation}  
with $m= 4 t_\perp/ 3 t^2 a^2$ being the mass of the parabolic dispersion. In bilayer graphene $m \approx 0.03 m_e$, where $m_e$ is the mass of free electrons.
\\

\section{Effective low-energy theory}

\subsection{Lagrangian}
In the previous section, I have shown that near the Dirac points (as $q_x, q_y \rightarrow 0$), states in the fully gapped bands ($E_2 (k)$) are localized on the sites of A sublattices, dubbed as ``dimer sites". Hence, in the low-energy limit these sites are of no dynamical importance. In this section I will derive the form of the effective low-energy Hamiltonian after integrating out the high energy band, $E_2(k)$. Our derivation closely follows the one shown in Ref.~\onlinecite{vafek-RG} Nevertheless, it is worth reviewing that derivation briefly to facilitate further discussion. The partition function describing the free motion of the spinless fermions is
\begin{eqnarray}
&&{\cal Z}=e^{{-\int^\beta_0} d\tau \; \psi^*_B \; \left( \partial_\tau + H_{BB} \right)\; \psi_B} \times 
\int {\cal D} \psi^*_A {\cal D}\; \psi_A \;\nonumber \\
&&e^{{-\int^\beta_0} d\tau \; \psi^*_A \; \left( \partial_\tau + H_{AA} \right)\; \psi_A + \psi^*_A H_{AB} \psi_B + \psi^*_B H_{BA} \psi_A }, 
\end{eqnarray}
where $\psi_X \;=\; \left( \psi_{1,X}, \psi_{2,X} \right)$ with $X=A,B,$ are the two component spinors and $i=1,2$ correspond to the layer indices. For convenience, I here neglect the remote hopping amplitudes, e.g., \emph{trigonal warping} and keep only the inter and intralayer nearest-neighbor hopping terms. Then the two-dimensional matrices $H_{AB}$, $H_{BB}$, and $H_{AA}$ read as
\begin{equation}
H_{BB}=
\left( \begin{array}{cc}
0 & 0   \\
0 & 0  \\
\end{array} \right)
\; , \;
H_{AA}=
\left( \begin{array}{cc}
0 & t_\perp   \\
t_\perp & 0  \\
\end{array} \right),
\end{equation}  
and 
\begin{equation}
H_{AB}=
\left( \begin{array}{cc}
t \; f(k) & 0   \\
0 & t \; f^*(k)  \\
\end{array} \right)
\; , \;
H_{BA}=H^\dagger_{AB}.
\end{equation}
Even though, $H_{BB}$ is trivial at the bare level, once we integrate out the high energy modes, it gets renormalized. The single-particle Green's function for the $\psi_A$ field is
\begin{equation}
G_{AA} (i \omega_n)=\frac{i \omega_n I_2 + t_\perp \sigma_x}{\omega^2_n+t^2_\perp},
\end{equation}
where $\omega_n=(2 n+1) \pi T$ are the fermionic Matsubara frequencies, and $T$ is the temperature. Expanding the action to the quadratic order and integrating out the $\psi_A$ field, the renormalized partition function is obtained for the $\psi_{B}$ field as
\begin{eqnarray}
&&{\cal Z}_B=e^{-\frac{1}{\beta} \; \sum_{\omega_n} \int d\vec{x} {\cal L}_0 }, \nonumber \\
&=&e^{{-\frac{1}{\beta}}\; \sum_{\omega_n} \psi^*_B (i\omega_n) \; \left( -i \omega_n + H_{BB} -H_{AB} G_{AA} (i\omega_n) H_{BA}\right)\; \psi_B (i\omega_n)}, \nonumber \\
&=& e^{-\frac{1}{\beta} \; \sum_{\omega_n} \psi^*_B (i\omega_n) \; L_0  \psi_B (i\omega_n) }.
\end{eqnarray}
Since we are interested in the modes near the charge-neutrality points, one can simply set $\omega_n=0$ in the last term of the effective action.

\subsection{Hamiltonian}
The imaginary-time, non interacting Lagrangian ($L_0$) is related to the single-particle Hamiltonian ($H_0$) according to $L_0 = -i \omega_n + H_0$. Therefore, the low-energy Hamiltonian, to the quadratic order in momentum is  
\begin{equation}
H_0= 
\left( \begin{array}{cc}
0 & \frac{v^2_F}{t_\perp} (q_x+iq_y)^2 \\
\frac{v^2_F}{t_\perp} (q_x- iq_y)^2 & 0 \\
\end{array} \right),
\end{equation}
near one of the Dirac point at $\vec{K}$. Taking into account the Fourier modes near the other Dirac point at $-\vec{K}$, the four dimensional Hamiltonian describing the low excitations (spinless) reads as 
\begin{equation}
H_0 = \gamma_2 \; \left(\frac{q^2_x-q^2_y}{2 \; m} \right)\;+\; \gamma_1 \; \left( \frac{- 2 \;q_x \;q_y}{2 \; m} \right),
\label{hamilfree}
\end{equation}
in the basis of the four-component spinor $\Psi(\vec{x})$, defined as 
\begin{eqnarray}
\Psi^\dagger(\vec{x})&=& \int^\Lambda \frac{d \;\vec{q}}{(2 \pi a)^2} e^{i \vec{q}\cdot \vec{x}} \; \bigg[ v^\dagger_1(\vec{K}+\vec{q}), \nonumber \\
&&v^\dagger_2(\vec{K}+\vec{q}),v^\dagger_1(-\vec{K}+\vec{q}),v^\dagger_2(-\vec{K}+\vec{q}) \bigg].
\label{psi-basic}
\end{eqnarray}
$\Lambda (\sim t^2_\perp/4 t \sim 200$ meV) is the high energy or the ultra violet cut off representing the range of energies over which the quasi particle dispersion is approximately parabolic. The mutually anti-commuting four component Hermitian $\gamma$ matrices belong to the representation \cite{herbut-juricic-roy}
\begin{equation}
{\gamma_0} =
 \left(\begin{array}{c c}
{\sigma_z} & 0 \\
0 & {\sigma_z}
\end{array}\right), \;
{\gamma_1} =
 \left(\begin{array}{c c}
{\sigma_y} & 0 \\
0 & -{\sigma_y}
\end{array}\right),\;
{\gamma_2} =
 \left(\begin{array}{c c}
{\sigma_x} & 0 \\
0 & {\sigma_x}
\end{array}\right).
\end{equation}
The two remaining anti-commuting matrices can then be chosen to be
\begin{equation}
{\gamma_3} =
 \left(\begin{array}{c c}
0 & {\sigma_y} \\
{\sigma_y} & 0
\end{array}\right),\quad
{\gamma_5} = \left(\begin{array}{c c}
0 & -i{\sigma_y} \\
i{\sigma_y} & 0
\end{array}\right).
\end{equation}
These matrices satisfy the anti-commuting Clifford algebra $\{ \gamma_\mu,\gamma_\nu \}= 2 \delta_{\mu \nu}$, for $\mu,\nu=0,1,2,3,5$. The form of the free Hamiltonian is identical to the one for monolayer graphene in quadratic order.\cite{karen-le-hur, herbut-freeHamiltonian} The underlying reason is as follows. After integrating out the high energy bands, the remaining lattice points on the $B$ sublattices also constitute a \emph{honeycomb lattice}, preserving the $C_{3v}$ symmetry around each site. Therefore, the free Hamiltonian needs to be invariant under a rotation by $2\pi/3$ around the Dirac points, which restricts the kinetic-energy Hamiltonian to the announced form.

\subsection{Symmetries}
The free Hamiltonian respects an emergent global chiral $SU_c(2)$ symmetry, generated by $\left\{ i \gamma_0 \gamma_3, i \gamma_0 \gamma_5, \gamma_{35} \right\}$, where $\gamma_{35}= i\gamma_3 \gamma_5$. The third entity of the group, is the generator of the translation.\cite{vafek-RG, herbut-juricic-roy} A similar chiral symmetry is also present in the emergent low-energy theory of the massless Dirac fermions in graphene,\cite{herbut-juricic-roy} and $d$-wave superconductors.\cite{d-wave} In addition to the chiral symmetry, $H_0$ is also invariant under the exchange of the layer indices, as well as the Dirac points.\cite{vafek-RG} These two reflection symmetries are generated by $I_{12}=\gamma_2$ and $I_K=i \gamma_1 \gamma_5$, when accompanied by the inversions of the momentum axis $q_y \rightarrow -q_y$ and $q_x \rightarrow -q_x$, respectively. Besides these, $H_0$ is also invariant under the time-reversal symmetry, since it describes motion of the free fermions on a lattice. The time-reversal symmetry is represented by an anti unitary operator $I_t= U\; K$, where $U$ is a unitary operator and $K$ is the complex conjugation. In our representation $U=i \gamma_1 \gamma_5=\sigma_1 \otimes I_2 \equiv I_K$. Therefore $I^2_t=+1$, as it should be, since $I_t$ is the time-reversal operator for the spinless fermions.\cite{gotfried} One can arrive at the same effective low-energy Hamiltonian by using the $\bf{K} \cdot p$ approach.\cite{vafek-RG, kpapproach}

\section{Spinless fermions in quantizing magnetic fields}

Before restoring the fermion's spin degrees of freedom, it is worth understanding the possible gapped states of the spinless fermions in bilayer graphene and their behavior in quantizing magnetic fields. In monolayer graphene, the linear dispersion makes all the short-ranged electron-electron interactions irrelevant near the noninteracting \emph{Gaussian} fixed point. The long-range Coulomb interaction ($\sim 1/r$) is also irrelevant, but only marginally.\cite{herbut-juricic-roy, juricic-conductivity, vozmediano} On the other hand, due to the quadratic band structure, all the short-ranged interactions are marginal in bilayer graphene. Hence, the semimetal-insulator transitions can take place even for weak interactions.\cite{vafek-kun, vafek-RG} Any order parameter, leading to a gap in the spectrum, must anticommute with the \emph{entire} free Hamiltonian, so that all the terms enter as a sum of the squares in the expression for energy. A finite gap then exists everywhere in the Brillouin zone. For spinless fermions in bilayer graphene, there are \emph{four} such candidates: $\left(\gamma_0\;,\;\gamma_3 \;,\;\gamma_5 \;,\; i\gamma_1 \gamma_2 \right)$. The first three members break the chiral $SU_c(2)$ symmetry of the free theory down to $U_c(1)$, whereas the last one lacks the time reversal symmetry, but preserves chiral symmetry. $\langle \Psi^\dagger \gamma_0 \Psi \rangle$ is the order parameter associated with the layer polarized state, leading to an imbalance of the electronic density among two layers. $\langle \Psi^\dagger i \gamma_1 \gamma_2 \Psi \rangle$ corresponds to Haldane's circulating current among the sites on the same sub-lattice.\cite{haldane} Otherwise, it propagates in the same directions in two layers, and the anomalous Hall state preserves the \emph{inversion} symmetry. The remaining two entities, $\langle \Psi^\dagger\left( \gamma_3, \gamma_5 \right) \Psi \rangle$ break the chiral as well as the time-reversal symmetry.\cite{chi-ken} One can, however, define an anti unitary operator as in Ref.~\onlinecite{chi-ken}, $\tilde{I}_t =i \gamma_{1} \gamma_{3}$ K, under which all three chiral symmetry-breaking masses are even, whereas the original time reversal odd mass, remains \emph{odd}. Otherwise, $\tilde{I}^2_t=-1$, and therefore does not corresponds to the true time-reversal operator. $\langle \Psi^\dagger\left( \gamma_3, \gamma_5 \right) \Psi \rangle$ correspond to Kekule currents, which additionally break the translational symmetry of the lattice into the Kekule pattern\cite{vafek-RG}. Otherwise the first (second) member is even (odd) under the exchange of two Dirac points. In monolayer graphene these two matrices are replaced by $i \gamma_0 \gamma_3$ and $i \gamma_0 \gamma_5$, giving rise to different realizations of the spatially modulated Kekule bond density waves.\cite{chamon-hou-mudry} However, they are time-reversal symmetric. Next I argue that even though the Kekule currents lack the time-reversal symmetry, only the Haldane's mass ($i \gamma_1 \gamma_2$) corresponds to the \emph{quantum anomalous Hall insulators}.

Placed in a weak quantizing magnetic field, bilayer graphene exhibits plateaus in Hall conductivity at fillings $\nu=\pm 4(n+1)$. The orbital effect of the magnetic field can be captured via a minimal substitution $q_i \rightarrow q_i -A_i$ in $H_0$ in Eq.\ (\ref{hamilfree}), giving $H_0[A]$. The magnetic field reads as $B= \epsilon_{ij} \partial_i A_j$. The spectrum of $H_0[A]$ is composed of a set of macroscopically degenerate Landau levels at well separated energies $E_n=\sqrt{n (n-1) B^2}$ with $n=0,1,2, \cdots$. The zeroth Landau level in bilayer graphene (with $n=0,1$) carries additional two fold orbital degeneracy due to the parabolic dispersion in the vicinity of the Dirac points\cite{falko, Kailasvuori}. States in the zeroth Landau level near two valleys ($\vec{K}$ and $-\vec{K}$) reside on the complementary layers, 1 and 2 respectively, for example. These two sets of zero energy state constitute a two-dimensional basis ${\cal H}_0$. Any matrix that commutes or anti commutes with the Hamiltonian $H_0[A]$ leaves that space invariant. There are \emph{four} matrices, falling into the second category, $\{ \gamma_0, \gamma_3, \gamma_5, i\gamma_1 \gamma_2 \}$. Together they also close a $Cl(3)\times U(1)$ algebra of the order parameters, where the $U(1)$ part is constituted by the last entry.\cite{cl3} To understand the behavior of these orders in the presence of magnetic fields, let us consider an auxiliary Hamiltonian
\begin{eqnarray}
H[\mathbf{m}] = H_0[A]+m_1 \gamma_0 + m_2 \gamma_3 + m_3 \gamma_5 + m_4 i \gamma_1 \gamma_2.
\end{eqnarray}    
The eigenvalues of $H[m_1,m_2,m_3,0]$ are at $\pm \sqrt{n (n-1) B^2 + m^2_1 + m^2_2 + m^2_3}$. Any linear combination of $m_1, m_2, m_3$ reduces the chiral $SU_c(2)$ symmetry to a $U_c(1)$. Therefore, in the presence of the chiral symmetry breaking orders the zeroth Landau level splits to $E_0 = \pm \sqrt{m^2_1 + m^2_2 + m^2_3}$, whereas Landau levels at finite energies are only \emph{shifted}. Hence it is \emph{always} energetically advantageous for the system to develop such \emph{mass} orders in the presence of the magnetic fields, to maximally lower the energy. The mechanism of developing a chiral symmetry breaking \emph{mass} order is known as ``magnetic catalysis", discussed previously in the context of Dirac fermions subject to magnetic fields.\cite{gusynin, reviewQHEgraphene} It can also be the underlying mechanism behind the formation of Hall states at fillings $\nu=0, \pm 1$ in monolayer graphene.\cite{catalysis-monolayer} However the energy spectrum of $H[0,0,0,m_4]$ within the zeroth Landau level is sign($m_4$)$m_4$. The rest of the Landau levels are shifted to $\pm \sqrt{n(n-1)B^2 +m^2_4}$. A finite $m_4$ therefore shifts the entire zeroth Landau level, and can only be realized by changing the \emph{filling factor} from the neutrality ($\nu=0$). Concomitantly, it leads to quantized Hall conductivity $\sigma_{xy}= \pm 2 \frac{e^2}{h}$. Even though $m_3$ and $m_4$ break the time-reversal symmetry, they yield zero Hall conductivity. Therefore, only the Haldane order ($i \gamma_1 \gamma_2$) corresponds to the quantum anomalous Hall insulator. A similar conclusion can be arrived at upon computing the expectation values of the aforementioned bi linears, with the non interacting wave-functions of the zeroth Landau level.\cite{rahul-levitov-QAH} It can also be justified from that fact that $\vec{M}_{c,V}=\left( \gamma_0\;,\;\gamma_3 \;,\;\gamma_5 \right)$ transform as a vector under the chiral $SU_c(2)$ rotation generated by $\{ i \gamma_0 \gamma_3, i \gamma_0 \gamma_5, i \gamma_3 \gamma_5 \}$, whereas $M_{c,S}= i \gamma_1 \gamma_2$ is  a scalar under the chiral transformation. Hence all the three chiral symmetry-breaking orders must lead to an identical spectrum of Hall conductivity in quantizing magnetic fields, $\nu=0$ Hall state.

Recently, there have been various proposals for the electronic ground state in bilayer graphene subject to magnetic field.\cite{genrefinBLGMag} In particular, a recent experiment shows that bilayer graphene exists in a gapped phase in the absence of magnetic field when the system is dual gated.\cite{Lau} Otherwise the gap increases monotonically with the magnetic field, whereas it gradually disappears upon applying a weak perpendicular electric field. These observations predict that the zero field order is likely to be the layer antiferromagnet and rules out the possibility of the layer polarized state. However, for weak fields the most promising candidate of the ordered state is possibly a partially spin polarized state, with coexistence of an easy plane (perpendicular to the applied magnetic field) antiferromagnet and an easy axis (in the direction of the field) magnetization.\cite{roybilayerscaling} 

\section{Nambu-Dirac fermion}

In the last section, I considered the insulating orders for spinless fermions in bilayer graphene. I name them as ``mass orders". Next, we wish to find all the ordered states, including the insulators, superconductors, semi-metals, and the gapless superconductors. To accommodate all the order parameters, we need to introduce a particle-hole doubled 16-component Nambu-Dirac fermion, defined as $\Psi=\left(\Psi_p,\Psi_h \right)^\top$, with $\Psi_p=\left( \Psi_{p,\uparrow}, \Psi_{p,\downarrow}\right)^\top$, and $\Psi_p=\left( \Psi_{h,\downarrow}, - \Psi_{h,\uparrow}\right)^\top$ with 
\begin{eqnarray}
&& \Psi^\top _{p \sigma}  (\vec{q}) = \nonumber \\
&&\bigg[ v _{1,\sigma} (\vec{K} + \vec{q}), v _{2,\sigma} (\vec{K}+ \vec{q}), v_{1,\sigma} (-\vec{K}+ \vec{q}),v_{2,\sigma} (-\vec{K}+ \vec{q}) \bigg], \nonumber \\
\end{eqnarray}
\begin{eqnarray}
&& \Psi^\top _{h \sigma}  (\vec{q}) = \nonumber \\
&&\bigg[ v^\dagger_{2,\sigma} ( \vec{K} - \vec{q}), v^\dagger_{1,\sigma} (\vec{K} - \vec{q}), v^\dagger_{2,\sigma} (-\vec{K}- \vec{q}), v^\dagger_{1,\sigma} (-\vec{K}- \vec{q}) \bigg], \nonumber \\
\end{eqnarray}
similar to the one recently considered in the context of mono-layer graphene. \cite{kekuleSC} In this basis, the tight-binding Hamiltonian in the low-energy approximation then takes the form
\begin{equation}
H_t \;=\; \sum_{\vec{q}} \Psi^\dagger (\vec{q}) H_0 \Psi (\vec{q}), 
\end{equation}  
where $H_0$ in the first quantization reads as
\begin{equation}
H_0 \;=\; \tau_3 \otimes \sigma_0 \otimes \left[ \gamma_2 \; \left(\frac{q^2_x-q^2_y}{2 \; m} \right)\;+\; \gamma_1 \; \left( \frac{- 2 \;q_x \;q_y}{2 \; m} \right)  \right], 
\label{kinetichamil16}
\end{equation}
where the four-dimensional $\gamma$ matrices belong to the aforementioned representation. The effective mass of the quasiparticle excitations is $m=t_\perp/v^2_F$, where $v_F=\sqrt{3}\;t a/2$ is the Fermi velocity in single layer graphene. The two component Pauli matrices $(\tau_0,\vec{\tau})$ operates on Nambu'sspace, whereas $(\sigma_0,\vec{\sigma})$ on the spin indices. Before we proceed to identify the massive and the gapless order parameters, it is worth pausing to register the symmetries of the quadratic Hamiltonian $H_0$. The reflection symmetries of $H_0$ under the exchange of the layers and the Dirac points, mentioned in Sec. II, are respectively generated by $I_{12}=\tau_0 \otimes \sigma_0 \otimes \gamma_2$ and $I_K= \tau_0 \otimes \sigma_0 \otimes i \gamma_1 \gamma_5$. The generator of translation is $P=\tau_3 \otimes \sigma_0 \otimes i \gamma_3 \gamma_5$. $H_0$ also commutes with the number operator $N=\tau_3 \otimes \sigma_0 \otimes I_4$. One advantage of this representation of the Nambu-Dirac spinor is that the three generators of the rotation of the electron's spin assume a simple form $\vec{S}=\tau_0 \otimes \vec{\sigma} \otimes I_4$. The free Hamiltonian ($H_0$) commutes with $\vec{S}$.

Besides the above 16-component Nambu-Dirac spinor, I also define an unitarily equivalent one, $\Phi=(\Phi_p,\Phi_h)^\top$, where $\Phi_p=(\Phi_{p,\uparrow},\Phi_{p,\downarrow})^\top$ and $\Phi_h=(\Phi_{h,\uparrow},\Phi_{h,\downarrow})^\top$, with
\begin{equation}
\Phi _{p \sigma}  (\vec{q}) \;=\; \Psi^\top _{h \sigma}  (- \vec{q}),
\end{equation}
and
\begin{eqnarray}
&& \Phi^\top _{h \sigma}  (- \vec{q}) = \nonumber \\
&&\bigg[ v^\dagger_{1,\sigma} ( \vec{K} - \vec{q}), v^\dagger_{2,\sigma} (\vec{K} - \vec{q}), v^\dagger_{1,\sigma} (-\vec{K}- \vec{q}), v^\dagger_{2,\sigma} (-\vec{K}- \vec{q}) \bigg]. \nonumber \\
\end{eqnarray}
This representation ($\Phi$) is related to the previous one ($\Psi$) according to 
\begin{equation}
\Phi (\vec{q}) \;=\; \bigg[ I_8 \; \oplus \; \left( i \sigma_2 \otimes \gamma_2 \right) \bigg] \Psi(\vec{q}),
\end{equation}
where $I_8$ is the eight-dimensional unity matrix. Soon we will appreciate the usefulness of such unitary equivalence. A similar definition of spinor has recently been used to describe all the possible masses in monolayer graphene.\cite{chamon-masses} We can immediately notice that there are altogether 64 matrices (${\cal M}$), where
\begin{equation}
{\cal M} = \left(\tau_0, \tau_3 \right) \otimes \left(\sigma_0, \vec{\sigma} \right)  \otimes \left(\gamma_0,\gamma_3,\gamma_5,i\gamma_1 \gamma_2 \right),
\end{equation}
and 
\begin{equation}
{\cal M} = \left(\tau_1, \tau_2 \right) \otimes \left(\sigma_0, \vec{\sigma} \right)  \otimes \left(I_4, i \gamma_0 \gamma_3, i\gamma_0 \gamma_5,i\gamma_3 \gamma_5 \right),
\end{equation}
which anticommute with the kinetic-energy Hamiltonian $H_0$ in Eq.\ (\ref{kinetichamil16}). Naively, one may therefore expect that all 64 fermionic bilinears of the form $\Psi^\dagger {\cal M} \Psi$ will lead to a gapped quasiparticle spectrum. However, such abundance is clearly an artifact of the Nambu's doubling of the original degrees of freedom. Next I show that this number is drastically reduced by some algebraic constraints that the mass order parameters need to satisfy. Upon imposing the constraints, I show that there are only 28 bilinears, which lead to a gap in the excitation spectrum.

\section{Mass orders}

Next we wish to derive the algebraic constraints that all the mass order parameters need to satisfy. We derive them separately for insulating and superconducting orders.

\subsection{Insulators}

All the insulating order parameters commute with the number operator ($N$). Therefore, one can have either $\tau_0$ or $\tau_3$ in the Nambu space. Hence, in general, all the insulating order (INS) is restricted to the following form:
\begin{equation}
\mbox{INS}\;=\;\Psi^\dagger \; \left[ \begin{array}{c|c}
{M_1} & {0}  \\
\hline {0} & {M_2}
\end{array} 
\right] \; \Psi,
\end{equation}  
where $M_1$ and $M_2$ are eight-dimensional Hermitian matrices. Let us first consider the insulating orders with $\tau_0$ in Nambu space. Then $M_1\;=\;M_2$ and 
\begin{eqnarray}
\mbox{INS}&=&
\Phi^\dagger \; \left\{ 
\left( \begin{array}{c|c}
{I_8} & {0}  \\
\hline {0} & {i R}
\end{array} \right) 
\left(
\begin{array}{c|c}
{M_1} & {0}  \\
\hline {0} & {M_1}
\end{array} \right)
\left(
\begin{array}{c|c}
{I_8} & {0}  \\
\hline {0} & {- i R}
\end{array} \right)
\right\} \; \Phi  \nonumber \\
&=& \Phi^\dagger_p M_1 \Phi_p + \Phi^\dagger_h \bigg[ R \times M_1 \times R \bigg] \Phi_h \nonumber \\
&=& \Phi^\dagger_p \left( M_1 - \left( R \times M_1 \times R \right)^\top \right)\Phi_p, \nonumber
\end{eqnarray}
where $R=\sigma_2 \otimes \gamma_2$. In our representation, all the matrices are either \emph{purely} real or \emph{purely} imaginary, and hence, either $M^\top_1=+ M_1$ or $M^\top_1=- M_1$, respectively. Together with this condition, $\left( \sigma_2 \otimes \gamma_2\right)^\top=-\sigma_2 \otimes \gamma_2$ gives us the requisite condition for the nonzero expectation value of the gapped insulating orders ($\langle \mbox{INS}\rangle\neq 0$),    
\begin{equation}
\left( \sigma_2 \otimes \gamma_2 \right) \times M^\top_1 \times \left( \sigma_2 \otimes \gamma_2\right) \; = \; - M_1.
\label{tau0INS}
\end{equation}
Only the following bilinears meet the above criteria, \\ 
\begin{enumerate}
\item{$\langle \Psi^\dagger \left( \tau_0 \otimes \sigma_0 \otimes \gamma_0 \right) \Psi \rangle=\Delta_{LP}:$ layer polarized state,}
\item{$\langle \Psi^\dagger \left( \tau_0 \otimes \sigma_0 \otimes i \gamma_1 \gamma_2 \right) \Psi \rangle=\Delta_{AH}:$ anomalous Hall insulator,}
\item $\langle \Psi^\dagger \left( \tau_0 \otimes \sigma_0 \otimes  \gamma_5 \right) \Psi \rangle=\Delta^{0}_{odd} :$ \emph{odd} spin singlet Kekule current ,
\item $\langle \Psi^\dagger \left( \tau_0 \otimes \vec{\sigma} \otimes  \gamma_3 \right) \Psi \rangle=\vec{\Delta}_{even} : \;$ \emph{even} spin-triplet Kekule current.
\end{enumerate}
The classification \emph{even} and \emph{odd} reflect the transformation of the Kekule order parameters under the exchange of the Dirac points. On the other hand, with $\tau_3$ in the Nambu space, i.e. $M_1=-M_2$, the condition in Eq.~(\ref{tau0INS}) reverts to 
\begin{equation}
\left( \sigma_2 \otimes \gamma_2 \right) \times M^\top_1 \times \left( \sigma_2 \otimes \gamma_2 \right) \; = \; + M_1.
\label{tau3INS}
\end{equation}
It allows us to capture the remaining insulating masses: \\
\begin{enumerate}
\item[5.] $\langle \Psi \left( \tau_3 \otimes \sigma_0 \otimes \gamma_3 \right) \Psi \rangle=\Delta^{0}_{even}:$ \emph{even} spin-singlet Kekule current,
\item[6.] $\langle \Psi \left( \tau_3 \otimes \vec{\sigma} \otimes \gamma_0 \right) \Psi \rangle=\vec{\Delta}_{LAF}:$ layer antiferromagnet,
\item[7.] $\langle \Psi \left( \tau_3 \otimes \vec{\sigma} \otimes i \gamma_1 \gamma_2 \right) \Psi \rangle=\vec{\Delta}_{SH}:$ spin Hall insulator
\item[8.] $\langle \Psi \left( \tau_3 \otimes \vec{\sigma} \otimes \gamma_5 \right) \Psi \rangle =\vec{\Delta}_{odd}: \;$ \emph{odd} spin-triplet Kekule current.
\end{enumerate}
Hence, there are eight-insulating orders, towards which the semimetallic bilayer graphene may suffer instabilities and open gap at the Dirac points. As one can see, it requires 16 linearly independent matrices, anticommuting with the kinetic-energy Hamiltonian, to define all the gapped insulating orders. However, some of them are connected by the ordinary rotations of electrons spin. All the spin-singlet orders, e.g., layer polarized, quantum anomalous, singlet Kekule currents, break the discrete layer-inversion symmetry, generated by $I_{12}$. On the other hand, the spin-triplet orders, e.g., the layer antiferromagnet, the quantum spin Hall insulator, and the two spin Kekule orders, additionally break the $SU(2)$ spin rotational symmetry. Thus, such ordered phases are always accompanied by two massless \emph{Goldstone modes}. In the low-energy limit, there is an internal $U(1)$ symmetry among various realizations of Kekule orders (for both spin singlet and triplet). However, such an \emph{emergent} internal symmetry is clearly an artifact of the parabolic band approximation of the dispersion near the Dirac points. The underlying honeycomb lattice always reduces such an emergent $U(1)$ symmetry to a discrete $C_{3v}$ symmetry. Hence, in a strict sense, there is no Goldstone mode in the Kekule phase.

\subsection{Gapped superconductors}

Next we find all the gapped superconducting states available for fermions in bilayer graphene to pair into. Superconducting or the \emph{off-diagonal} order parameters, on the other hand \emph{anti commute} with the number operators ($N$) and takes the following form
\begin{equation}
\mbox{SC}\;=\;\Psi^\dagger \left[ \begin{array}{c|c}
{0} & {M}  \\
\hline {M^\dagger} & {0}
\end{array} 
\right] \Psi,
\end{equation}
in general. $M$ is an eight-dimensional matrix. SC can also be cast in the following form:
\begin{eqnarray}
\mbox{SC}&=&
\Phi^\dagger \left\{ 
\left( \begin{array}{c|c}
{I_8} & {0}  \\
\hline {0} & {i R}
\end{array}
\right) 
\left( \begin{array}{c|c}
{0} & {M}  \\
\hline {M^\dagger} & {0}
\end{array} \right) 
\left( \begin{array}{c|c}
{I_8} & {0}  \\
\hline {0} & {- i R}
\end{array} \right)
\right\} \Phi \nonumber \\
&=&\Phi^\dagger \bigg[
\begin{array}{c|c}
{0} & {- i M \;R}  \\
\hline {i \;R \; M^\dagger} & {0}
\end{array} 
\bigg]\; \Phi.
\end{eqnarray}
Therefore, any off-diagonal order with nontrivial expectation value must satisfy the following condition:
\begin{equation}
\left( \sigma_2 \otimes \gamma_2 \right) \; M^\top \;=\; M \; \left( \sigma_2 \otimes \gamma_2 \right).
\label{constrainSC}
\end{equation}
The superconducting orders that concur with this constraint are
\begin{enumerate}
\item $\langle \Psi \left[ ( \tau_1 \cos \phi  + \tau_2 \sin \phi )  \otimes \sigma_0 \otimes i\gamma_0 \gamma_3 \right] \Psi \rangle=\Delta_{s}: $ spin-singlet $s$-wave,
\item $\langle \Psi \left[ ( \tau_1 \cos \phi  + \tau_2 \sin \phi )   \otimes \vec{\sigma} \otimes i\gamma_0 \gamma_5 \right] \Psi \rangle$ $=\vec{\Delta}_{f}$ : spin-triplet $f$-wave,
\item $\langle \Psi \left[ ( \tau_1 \cos \phi  + \tau_2 \sin \phi )  \otimes \sigma_0 \otimes I_4 \right] \Psi \rangle$ $= \Delta^{Kek}_{s}:$ spin-singlet $s$-Kekule, 
\item $\langle \Psi \left[ ( \tau_1 \cos \phi  + \tau_2 \sin \phi )  \otimes \sigma_0 \otimes i \gamma_3 \gamma_5 \right] \Psi \rangle$ $=\Delta^{Kek}_{p}: \;$ spin-singlet $p$-Kekule,
\end{enumerate} 
superconductors. Here $\phi$ is the superconducting phase. The s-wave order is even under the exchange of layers and Dirac points and translationally invariant. A similar s-wave phase can also be realized in neutral mono-layer graphene if the onsite attractive interaction is sufficiently strong.\cite{zhao} The translationally invariant $f$-wave SC order, is odd (even) under the Dirac point (layer) exchange, but changes its sign six times around the Brillouin zone, similar to the one in mono-layer graphene.\cite{honerkamp} The Kekule superconductors, on the other hand, break the translational symmetry of the lattice into Kekule pattern, odd under the exchange of layers and spin singlet. Otherwise, $s$($p$)-Kekule is even (odd) under the exchange of two Dirac points. In a monolayer graphene, however, the pertinent gapped Kekule superconductors are spin-triplet.\cite{kekuleSC}

Therefore, fermions in bilayer graphene can pair into four gapped superconducting phases. However, one requires 12 linearly independent matrices to define all of them. Hence, altogether there are 
\begin{center}
16 (insulators) + 12 (superconductors) = 28
\end{center}         
fermionic bi linears which can give rise to a gap in the quasiparticle dispersion, near the parabolic degeneracy points.

\section{Gapless states}

Besides the fully gapped states, electron-electron interactions in bilayer graphene can also support various gapless phases\cite{vafek-kun}. Before we proceed to recognize all possible gapless states, it is worth pausing to provide a generalized definition of such a phase, appropriate for bilayer graphene.

\subsection{Definition}
For simplicity let us consider spinless fermions once again. Recall that in the vicinity of the parabolic touching points the kinetic-energy Hamiltonian takes the form
\begin{equation}
H_0 [q_x,q_y] \:=\: \gamma_2 \; \left( \frac{q^2_x-q^2_y}{2m} \right) \:+\: \gamma_1 \; \left( \frac{-2 q_x q_y}{2m} \right). 
\end{equation}
\emph{A gapless order is associated with a matrix (${\cal M}$) which anticommutes with only one of the matrices appearing in the free Hamiltonian (namely, $\gamma_1$ and $\gamma_2$), while it commutes with the other one.} If such an order develops a finite expectation value then the parabolic bands split into two Dirac cones, separated by an amount proportional to the size of the gapless order. Otherwise, the gapless orders are invariant under rotation by $\pi$. Note that the parabolic bands are associated with a Berry's phase $2 \pi$. Therefore, an order parameter can in principle split the parabolic band touchings into two Dirac cones, which carry a Berry's phase $\pi$. To gain further intuition about the gapless orders let us consider \emph{two} specific examples: (1) $M\;=\; \Delta_1 \gamma_1$ and (2) $M\;=\; \Delta_2 \gamma_2$. Here $\Delta_{1,2}$ correspond to sizes of the two gapless orders. In the former situation, the Dirac cones appear at $q_x \;=\; q_y \;=\; \pm \sqrt{ m \Delta_1}/2m$, whereas a finite $\Delta_2$, gives birth to two Dirac cones at $q_x\;=\;0, q_y \;=\; \pm \sqrt{2 \Delta_2 m}/2m$. These two gapless orders correspond to \emph{nematic} orders. Below I present several other examples of the gapless states.

\subsection{Semimetals}

Next, we consider the 16-component Nambu-Dirac spinor ($\Psi$) and wish to write down all the possible semimetals (commute with the number operator) as well as gapless superconductors (anticommute with the number operator). Naively, one can assume there are all together 64 semimetallic orders parameters, of the form $\left(\tau_0,\tau_3 \right) \otimes \left(\sigma_0,\vec{\sigma} \right) \otimes Nm$, where $Nm \in \left( \gamma_1,\gamma_2,i\gamma_0 \gamma_1,i\gamma_0 \gamma_2,i\gamma_1 \gamma_3, i\gamma_2 \gamma_3, i\gamma_1 \gamma_5,i\gamma_2 \gamma_5 \right)$. However, to acquire finite expectation values, the semi-metallic orders need to satisfy one of the constraints Eqs.\ (\ref{tau0INS}) or \ (\ref{tau3INS}). In conjunction with the above definition, these constraints yield the following spin-singlet semimetals: 
\begin{enumerate}
\item $\langle \Psi^\dagger \left( \tau_0 \otimes \sigma_0 \otimes \gamma_1, \tau_0 \otimes \sigma_0 \otimes \gamma_2 \right) \Psi \rangle =\vec{\Delta}_{12}$,
\item $\langle \Psi^\dagger \left( \tau_3 \otimes \sigma_0 \otimes i \gamma_0 \gamma_2, \tau_3 \otimes \sigma_0 \otimes i \gamma_0 \gamma_1 \right) \Psi \rangle=\vec{\Delta}^0_{12}$,
\item $\langle \Psi^\dagger \left( \tau_0 \otimes \sigma_0 \otimes i \gamma_2 \gamma_3, \tau_0 \otimes \sigma_0 \otimes i \gamma_1 \gamma_3 \right)\Psi \rangle=\vec{\Delta}^3_{12}$,
\item $\langle \Psi^\dagger \left( \tau_3 \otimes \sigma_0 \otimes i \gamma_2 \gamma_5, \tau_3 \otimes \sigma_0 \otimes i \gamma_1 \gamma_5 \right) \Psi \rangle=\vec{\Delta}^5_{12}$.
\end{enumerate}
$\vec{\Delta}_{12}, \vec{\Delta}^0_{12}$ correspond to nematic order, and $\vec{\Delta}^3_{12}, \vec{\Delta}^5_{12}$ to a \emph{charge density wave}. Elements within the same group are connected by rotation of $\pi/2$ around the Dirac points, generated by $I_{RK}=\tau_0 \otimes \sigma_0 \otimes i \gamma_1 \gamma_2$. The first member of each group is odd under exchanges of two layers, whereas the second entry is even under the same operation. The gapless orders $\vec{\Delta}^3_{12}$ and $\vec{\Delta}^5_{12}$ break the translational symmetry, whereas $\vec{\Delta}_{12}$ and $\vec{\Delta}^0_{12}$ are invariant translation invariant. In the analogy with the terminology of \emph{liquid crystals}, $\vec{\Delta}^3_{12}$ and $\vec{\Delta}^5_{12}$ can be named as \emph{smectic phases}\cite{chaikin-lubensky}. Under the exchange of two Dirac points, the first member of $\vec{\Delta}_{12}$, and the second members of $\vec{\Delta}^0_{12}$, $\vec{\Delta}^3_{12}$, $\vec{\Delta}^5_{12}$, are even, whereas the remaining members are odd. One can also write down all the bilinears (order parameters) corresponding to the \emph{triplet-semimetals},    
\begin{enumerate}
\item $ \langle \Psi^\dagger \left( \tau_3 \otimes \vec{\sigma} \otimes \gamma_1, \tau_3 \otimes \vec{\sigma} \otimes \gamma_2 \right)\Psi \rangle=\vec{\Delta}_{12,t}$,
\item $\langle \Psi^\dagger \left( \tau_0 \otimes \vec{\sigma} \otimes i \gamma_0 \gamma_2, \tau_0 \otimes \vec{\sigma} \otimes i \gamma_0 \gamma_1 \right)\Psi \rangle=\vec{\Delta}^0_{12,t}$,
\item $\langle \Psi^\dagger \left( \tau_3 \otimes \vec{\sigma} \otimes i \gamma_2 \gamma_3, \tau_3 \otimes \vec{\sigma} \otimes i \gamma_1 \gamma_3 \right)\Psi \rangle=\vec{\Delta}^3_{12,t}$,
\item $\langle \Psi^\dagger \left( \tau_0 \otimes \vec{\sigma} \otimes i \gamma_2 \gamma_5, \tau_0 \otimes \vec{\sigma} \otimes i \gamma_1 \gamma_5 \right)\Psi \rangle=\vec{\Delta}^5_{12,t}$.
\end{enumerate}
Transformations of the triplet-gapless orders under the exchange of two layers, Dirac points, translational are identical to the ones for spin-singlet gapless orders, since all the symmetry operators bear $\sigma_0$ (two-dimensional identity matrix) in the spin sectors. Apart from the massless Dirac fermionic excitations, triplet-gapless phases are also accompanied by two massless Goldstone modes, arising from the spontaneous breaking of spin rotational symmetry. In our notation $\vec{\Delta}_{12,t}$, $\vec{\Delta}^0_{12,t}$ correspond to \emph{spin-nematic} orders, and $\vec{\Delta}^3_{12,t}$, $\vec{\Delta}^5_{12,t}$ to a \emph{spin-density wave}.

\subsection{Gapless superconductors}

Fermions in bilayer graphene can also pair into various gapless superconducting states. Though, one can once again find 64 Nambu-Dirac bilinears, which anticommutes with the number operator and one of the matrices in the free Hamiltonian, only the following pairing order parameters satisfy Eq.\ (\ref{constrainSC}): 
\begin{enumerate}
\item $\langle \Psi^\dagger \left[( \tau_1 \cos \phi  + \tau_2 \sin \phi )  \otimes \sigma_0 \otimes \gamma_1 \right] \Psi \rangle=\Delta_1$,
\item $\langle \Psi^\dagger \left[( \tau_1 \cos \phi  + \tau_2 \sin \phi )  \otimes \sigma_0 \otimes \gamma_2 \right] \Psi \rangle=\Delta_2$,
\item $\langle \Psi^\dagger \left[( \tau_1 \cos \phi  + \tau_2 \sin \phi )  \otimes \sigma_0 \otimes i \gamma_0 \gamma_1 \right] \Psi \rangle=\Delta_{01}$,
\item $\langle \Psi^\dagger \left[( \tau_1 \cos \phi  + \tau_2 \sin \phi )  \otimes \sigma_0 \otimes i \gamma_0 \gamma_2 \right] \Psi \rangle=\Delta_{02}$,
\item $\langle \Psi^\dagger \left[( \tau_1 \cos \phi  + \tau_2 \sin \phi )  \otimes \vec{\sigma} \otimes i \gamma_1 \gamma_3 \right] \Psi \rangle=\vec{\Delta}_{13}$,
\item $\langle \Psi^\dagger \left[( \tau_1 \cos \phi  + \tau_2 \sin \phi )  \otimes \vec{\sigma} \otimes i \gamma_2 \gamma_3 \right] \Psi \rangle=\vec{\Delta}_{23}$,
\item $\langle \Psi^\dagger \left[( \tau_1 \cos \phi  + \tau_2 \sin \phi )  \otimes \sigma_0 \otimes i \gamma_1 \gamma_5 \right] \Psi \rangle={\Delta}_{15}$,
\item $\langle \Psi^\dagger \left[( \tau_1 \cos \phi  + \tau_2 \sin \phi )  \otimes \sigma_0 \otimes i \gamma_2 \gamma_5\right] \Psi \rangle=\Delta_{25}$,
\end{enumerate}
where $\phi$ is the superconducting phase. The last four superconducting orders preserve the translational symmetry and the Cooper pairs are formed by pairing fermions with momentum $\vec{K}+\vec{q}$ and $-\vec{K}-\vec{q}$ or $\vec{q} \rightarrow - \vec{q}$. The remaining four pairings are spatially inhomogeneous with periodicity $2 \vec{K}$ and break the translational symmetry. Cooper pairs in those channels are formed by gluing the fermions with momenta $\vec{K}+\vec{q}$, $\vec{K}-\vec{q}$ and $\vec{K} \rightarrow -\vec{K}$. I name them as ``gapless-Fulde-Ferrell-Larkin-Ovchinikov" superconductors. $\Delta_1$, $\Delta_{02}$, $\vec{\Delta}_{23}$, ${\Delta}_{25}$ are even, while the remaining four pairings are odd, under the exchange of two layers. On the other hand, $\Delta_1$, $\Delta_{01}$, $\vec{\Delta}_{13}$, $\Delta_{25}$ change sign, while the remaining four pairings remain invariant under the exchange of two Dirac points. Elements from each groups $\left(\Delta_1,\Delta_2\right)$, $\left(\Delta_{01},\Delta_{02}\right)$, $\left(\vec{\Delta}_{13}, \vec{\Delta}_{23}\right)$, and $\left(\Delta_{15},\Delta_{25}\right)$ transform into each other under the $\pi/2$ rotation around the Dirac points, generated by $I_{RK}$.

Above I present all the possible semi-metals, as well as gapless superconducting phases. All together there are 
\begin{center}
32 (semimetals) + 24 (superconductors)\; = \; 56
\end{center}
fermionic bi linears that define all the gapless orders in bilayer graphene. Note that one can write a set of matrices as $i H^{j}_0 {\cal M}_k$, where $H^{j}_0$ is one of the two matrices appearing in the kinetic energy Hamiltonian, whereas ${\cal M}_k$ is one of the 28 matrices, defining the massive order parameters. By construction, $i H^{j}_0 {\cal M}_k$ is Hermitian, anticommutes with one of the matrices in the free Hamiltonian, while it commutes with the other one, and hence meets the definition of the gapless order parameters. Therefore, the total number of semimetallic orders is $2 \; (\mathrm{for \; the \; index} \; j)  \times 16$ (number of insulators) $=32$, and that of the gapless superconductors is $2 \; (\mathrm{for \; the \; index} \; j) \times 12$ (number of gapped superconductors) $=24$, in accordance with our explicit computation, yielding total $56$ gapless order parameters. However, there are eight semimetals and eight gapless superconductors, as shown above.

\section{Interactions}  

In this section, I offer a qualitative discussion on the role of electron-electron interactions in bilayer graphene. As mentioned previously, all four fermion interactions are marginal in the bare level. That allows one to perform a weak-coupling expansion about the symmetric semimetallic ground state to study its instabilities towards the formation of various ordered states. The interacting theory in bilayer graphene can be expressed in terms of 18 quartic interactions.\cite{vafek-RG} However, not all 18 coupling constants are linearly independent. There exist a set of linear constraints, so-called ``Fierz identity", which allows one to write each of the quartic terms as a linear combination of the others.\cite{herbut-juricic-roy, itzykson} Such linear constraints restrict the number of independent quartic terms to 9. For example, one can write all the interactions in the spin-triplet channel as linear combinations of the ones in the singlet channel.

With repulsive Hubbard interaction (U), a one-loop renormalization-group calculation shows that the system finds itself in a state with a staggered pattern of spin among the two layers, the layer anti-ferromagnet state. This prediction can also be justified from the strong-coupling physics.\cite{vafek-RG} It is expected that each of the layers is antiferromagnetically ordered at least when $t_\perp=0$ and $U/t \gg 1$.\cite{paiva} However, the relative orientation of the antiferromagnet order in two layers is arbitrary when $t_\perp=0$. Upon turning on $t_\perp$, the sublattice magnetization on two layers assumes a staggered pattern. Using a similar argument, one can also predict the possible ground state if the repulsion ($V_2$) among the fermions living on the next-neighbor sites on same layer is the strongest component of the finite-ranged Coulomb interaction. If the layers are completely decoupled ($t_\perp \equiv 0$), each layer is expected to find itself in the \emph{quantum spin Hall insulator} phase, at least when $V_2/t \gg 1$.\cite{raghu} This phase supports circulating currents among the sites of the same sublattice. Otherwise it orients in the opposite direction on two sublattices.\cite{haldane} Its orientation is opposite for two spin projections. The spin Hall insulator preserves the total time reversal symmetry as well as the inversion symmetry. The spin Hall insulator additionally breaks the spin rotational symmetry. Therefore the ordered phase is accompanied by one massive and two massless modes.\cite{raghu, herbut-roy-pseudo-catalysis} When $t_\perp =0$, the orientation of the Haldane's circulating current in two layers is completely independent. A small $t_\perp$, however locks the circulation in two layers in the same direction.

When the interaction is relatively long ranged possibly an unconventional phase, \emph{nematic order} arises\cite{vafek-kun, robert, robert-ginzburg, Lemonik}. Unlike the fully gapped phases, a nematic order splits the parabolic band into two Dirac cones. However, they appear at different locations in the Brillouin zone than the Dirac points, which, on the other hand, become gapped. The separation among these two cones is proportional to the magnitude of the nematic order. The experimentally observed gapless ordered state appears to be the nematic state, $\vec{\Delta}_{12}$.\cite{yacoby} Note that any lattice model with density-density interaction contains both intravalley (forward) as well as intervalley (back) scatterings. Their relative strength, however, depends on the range of the interaction. Therefore one can find a rich phase diagram of various correlated phases simply by tuning the relative strength of these two types of scatterings.\cite{robert-ginzburg, Lemonik}

If, on the other hand, the net interaction acquires an attractive component, fermions in bilayer graphene may condense into variety of superconducting states. An attractive interaction can arise, for example from electron-phonon interactions or a novel proximity effect. An on-site attraction can favor a spin-singlet $s$-wave superconducting order, as in monolayer graphene.\cite{zhao,robert-ginzburg} Attractive interaction among the fermions living on the same layer, but at the next-neighbor sites, can support a spin-triplet $f$-wave superconducting state. The superconducting order parameter changes its sign six times around the Brillouin zone, similar to the one appropriate for the monolayer graphene.\cite{honerkamp} Two spatially inhomogeneous, spin singlet superconductors may arise in bilayer graphene when electrons living on two layers attract each other. The order parameter is odd under the exchange of the layers, and breaks the translational symmetry of the honeycomb lattice. Otherwise the $s$- and $p$-Kekule states are, respectively, odd and even under the Dirac point exchange. At this moment, the microscopic origin of the gapless superconductors is unknown. However, in a recent work, the existence of some unconventional superconducting states has been proposed theoretically.\cite{Lemonik}

It is, however, admitted that the weak-coupling renormalization group analysis is biased towards the formation of the gapped states, at least when $T=0$, since the fully gapped states always maximally lower the energy of the ground state. However, only at finite temperature, where the free energy and entropy competes, this approach can capture the competition between the fully gapped and the gapless states, to a certain extent. Furthermore, the weak-coupling renormalization-group analysis tracks \emph{only} the leading instability around a scale, where the coupling constants, as well as the susceptibilities of several order parameters \emph{diverge} simultaneously. As shown in Ref.~\onlinecite{robert-ginzburg}, the non interacting ground state in bilayer graphene can destabilize towards the formation of several fully gapped states. Apart from several gapped states, only the gapless nematic state $\vec{\Delta}_{12}$ has been found at finite temperatures. Such outcomes possibly point towards the limitation of this technique, and demand other approaches, e.g., strong coupling, Monte Carlo\cite{Assad} studies of this problem, which can capture the possible appearance of several other interesting states, e.g., charge or spin density waves. 

\section{Experimental signatures}   

In this section I propose some simple experimental tools to determine the nature of the broken symmetry phases in bilayer graphene. Readers may consult Refs.~\onlinecite{nandkishore-classification} and \onlinecite{gorbar-optical}, where other experimental probes e.g. optical, magneto-optical effects, have been considered.

Broken symmetry phases, as mentioned above, can be classified into the following three broad categories: insulator, semimetal (nematic or smectic), superconductors. A clear distinction among these three classes of ordered states can be observed in the resistivity or minimal conductivity ($\sigma_{min}$) measurements. Below the transition temperature an insulating phase should discern a increasing resistivity or decreasing $\sigma_{min}$ with the temperature. Finally as $T \to 0$, $\sigma_{min} \to 0$\cite{velascoresistivity}. Below the superconducting transition temperature the resistivity should display a sharp drop to \emph{zero} (or to an extremely low value) \cite{tinkham}. However, any superconducting transition should be confirmed by observing the flux expulsion from the bulk of the system at sufficiently weak magnetic field (below $H_{c1}$) and temperature, the \emph{Meissner effect}. On the other hand, if the fermions in bilayer graphene condense into a semimetallic (gapless) state, $\sigma_{min}$ saturates to a \emph{finite value} as $T \to 0$, and across the transition the $\sigma_{min}$ typically displays a \emph{kink}, as found in Ref. \onlinecite{velascoresistivity}.

Previously, I have shown that there are many candidates for the insulating, semi-metallic, or superconducting ground states in bilayer graphene. After realizing to which class the broken-symmetry state falls into, one needs to perform a series of other experiment to pin down the exact nature of the ordered state. Let us first present distinct experimental signatures of various superconducting states, which can further be classified into two categories: fully gapped and gapless superconductors. These two types of pairings lead to different features in $dI/dV$ spectroscopy measurements. Any fully gapped state will show a zero signal in the spectroscopy measurement if $V<\Delta$(superconducting gap) at sufficiently low temperatures, while a sharp peak can be observed when $V \sim \Delta$. On the other had, gapless superconductors do not show any gapped structure in the spectroscopic measurements. Both the gapped or the gapless superconductors can be realized in spin-singlet or -triplet channels, and furthermore they can be spatially uniform or nonuniform (FFLO) in nature. Triplet superconductors are devoid of Pauli limiting field. $T_c$ for singlet paired states decreases in the presence of a weak parallel magnetic field, while that with an underlying triplet pairings remains unchanged\cite{tinkham}. A spatially scanned spectroscopy measurement, in principle, should discern periodic variation, with periodicity $2 \vec{K}$, if the underlying superconducting state is FFLO in nature, whereas that for the uniform state is expected to be insensitive to the location of measurement. The difference in the ground-state energy with various underlying fully gapped FFLO states, e.g., $\Delta^{kek}_s$, $\Delta^{kek}_p$ or any linear combination of these two states, is extremely tiny, and the difference arises only if we take into account the contribution from the states, residing far away from the Dirac points\cite{kekuleSC}. However, the quasiparticle excitations are not sharp far away from the charge neutrality point, and one can neglect their contribution to the free energy. Consequently, an internal $U(1)$ symmetry among various linear combinations of $\Delta^{kek}_s$ and $\Delta^{kek}_p$ emerges at low-energy, and distinction between these two pairings is irrelevant. Therefore, by performing a set of simple experiments, some of which I propose here, one can determine the nature of the underlying superconducting state in bilayer graphene.

Different insulating states also bear distinct experimental signatures. For example, an electric field, applied perpendicular to the bilayer graphene plane, either increases or decreases the gap of the layer polarized states, depending on its direction. The layer antiferromagnet order, on the other hand, decreases irrespective of the direction of the applied electric field. The hallmark signature of the anomalous Hall state is the quantization of off-diagonal conductivity $\sigma_{xy}=\pm 2 e^2/h$, in the absence of any applied magnetic field\cite{haldane}. The quantum spin Hall insulator, on the other hand, does not discern quantization of charge Hall response, but exhibits quantized spin Hall response\cite{kane-mele}. I have shown that various translational symmetry breaking orders, e.g. $\Delta^0_{even}$,$\Delta^0_{odd}$, $\vec{\Delta}^0_{even}$, $\vec{\Delta}^0_{odd}$ can gap out the quasi particle spectrum in the vicinity of the Dirac points. It is important to notice that $\Delta^0_{even}$, and $\Delta^0_{odd}$ are connected to each other by a chiral $U(1)$ symmetry. Although approximate, within the framework of emergent low-energy theory the chiral symmetry is a good symmetry, and if we neglect the contribution to the ground state energy from the states residing far from the charge neutrality point, these two states are energetically degenerate. Therefore, any linear combination of these two order parameter is energetically equally viable\cite{kekuleSC}. A similar conclusion can be made for the translational symmetry breaking spin-Kekule current orders $\vec{\Delta}^0_{even}$ and $\vec{\Delta}^0_{odd}$. I therefore do not wish to present any distinguishing feature among these two states. Otherwise, breaking of the translation symmetry by any order parameter, can be confirmed in a \emph{diffraction} experiment \cite{chaikin-lubensky}. The appearance of new peaks at sufficiently low temperatures (below the transition temperature) results from the breaking of translational symmetry, and the emergence of lattice structure with $2 \vec{K}$ periodicity. This tool can also be useful to distinguish various translational symmetry breaking gapless or smectic states, e.g. $\vec{\Delta}^{3}_{12}$, $\vec{\Delta}^{5}_{12}$, $\vec{\Delta}^{3}_{12,t}$, $\vec{\Delta}^{5}_{12,t}$, from the other gapless states, which preserve the translational symmetry (nematic)\cite{chaikin-lubensky}. The singlet and the triplet insulating states, which lack the same set of discrete symmetries, can be distinguished in \emph{specific-heat} measurement. Since the triplet state, additionally breaks the $SU(2)$ spin rotational symmetry, the ordered phase is accompanied by two \emph{Goldstone modes}. As a result, the specific heat will be finite in both the triplet ordered and symmetric semimetallic phase (due to the gapless fermions). If the the underlying state is spin-singlet, the specific heat should vanish as $T \to 0$.

Similar to the superconducting and the insulating orders, there are several viable candidates for the gapless or semi-metallic ground state in bilayer graphene. In the last paragraph, I have shown how one can separate the translation symmetry-breaking smectic and preserving nematic phases in a diffraction experiment. Otherwise, $\vec{\Delta}^{3}_{12}$ and $\vec{\Delta}^{5}_{12}$ together correspond to stripes or charge density wave order, whereas $\vec{\Delta}^{3}_{12,t}$ and $\vec{\Delta}^{5}_{12,t}$ correspond to spin density wave\cite{robert-ginzburg}, with periodicity $2 \vec{K}$. The spin structure of the density wave order can easily be detected from NMR experiments. For instance, upon applying a radio frequency (rf) signal some of the spins flip if the underlying state is the spin density wave. When the signal is then turned off, the flipped spin relaxes back to the ground-state configuration. The emitted rf signal is the signature of a spin density wave ordering. Otherwise, an anisotropic longitudinal conductivity is the characteristic feature of any gapless (nematic/smectic) state in bilayer graphene. One should note that two spin-singlet nematic orders $\vec{\Delta}_{12}$ and $\vec{\Delta}^{0}_{12}$ respectively preserve and break the time-reversal symmetry. Consequently, the latter order state can discern finite Hall conductivity even at zero magnetic field. However, as pointed out in Ref. \onlinecite{nandkishore-classification}, due to the intrinsic gapless nature of this state, the Hall conductivity will not be quantized.

\section{Summary and discussion}

To summarize, I here present all the possible ordered phases, including fully gapped massive as well as the gapless phases in bilayer graphene, and study their transformation under various symmetries (discrete and/or continuous). The parabolic bands touching each other at the Dirac points can be gapped out by spontaneously developing either eight insulating or four superconducting orders. On the other hand, fermions in bilayer graphene can also be realized in various gapless states. I here show that as all together eight semi-metallic and eight gapless superconducting states can be realized in bilayer graphene. The recently fabricated bilayer silicene \cite{bilayer-silicene-fabrication}, which shares a similar crystallographic structure as the bilayer graphene, also appears to be a promising ground to realize various ordered states. A first principle calculation predicts the possible appearance of chiral $d$-wave, $f$-wave superconductivity\cite{silicene-firstprinciple}. More recently, a $35$-meV superconducting gap has been reported in bilayer silicene \cite{silicene-SC}. In response to ongoing research activity in the field of bilayer graphene or silicene, our classification of all the possible low-energy ground states, and the proposals to detect the nature of the underlying broken symmetry states can provide valuable insights to search for novel unconventional states in these materials.

Here I have shown that the low-energy effective theory of gapless fermions can be described in terms of parabolic bands, touching each other at the Dirac points. However, upon taking into account the direct hopping amplitudes among the low-energy degrees of freedom ($B$ sites), such parabolic band touching splits into four Dirac cones. As a consequence, one still requires finite strength of interactions to stabilize various ordered phases. Such critical strength of interactions is expected to be much smaller than that for monolayer graphene, since the direct hopping is weak. It is also therefore quite interesting to study the nature of the \emph{quantum phase transitions} in bilayer graphene.\cite{QPT-monolayer}

\section{Acknowledgements}
The author is grateful to I. F. Herbut for many useful discussion and in particular for some constructive criticism on the Clifford algebraic (see the Appendix) structure of this problem. It is a pleasure to acknowledge O. Vafek, K. Yang, P. Goswami, L. Balicas, V. Czetkovic, R. E. Throckmorton for useful discussion. Author is in debt to A. S. Narayan and K. Yang for critical reading of the manuscript. Author was supported at National High Magnetic Field Laboratory by NSF Cooperative Agreement No.DMR-0654118, the State of Florida, and the U. S. Department of Energy. Hospitality of Les Houches Summer school on `Strongly interacting quantum systems out of equilibrium', and Aspen Center of Physics during the Winter Conference on Topological States of Matter, where part of the paper was prepared, is gratefully acknowledged.

\appendix

\section{Algebraic derivation of masses and nematic orders in bilayer graphene}

Here we present an alternate formulation to derive the number of massive and gapless order parameters in the bilayer graphene. In what follows next, our derivation is in similar spirit with the one for monolayer graphene, presented in Appendix A of Ref.~\onlinecite{herbut-cliffordalgebra}. However, the algebraic structures of these two problems enjoy significant differences, as I show below. Let us consider a 16 component Nambu-Dirac fermion $\Psi=\left( \Psi_p, \Psi_h \right)^\top$, so that the effective Hamiltonian describing the low-energy excitations reads as 
\begin{equation}
H_k=H_0 (\vec{k}) \oplus \left( - H^\top_0 (-\vec{k})\right),
\end{equation}   
where 
\begin{equation}
H_0 (\vec{k})= \sum_{i=1,2}\alpha_i \; d_i,
\end{equation}
with 
\begin{equation}
d_1=\frac{k^2_x-k^2_y}{2 m} \quad \mbox{and} \quad d_2=\frac{-2 k_x k_y}{2 m},
\end{equation}
$\alpha_1$ and $\alpha_2$ are eight-component Hermitian matrices.\cite{representation} Here we wish to find all the Hermitian matrices ($M_i$s) that anti commutes with Hamiltonian $H_k$, and develop gaps in the spectrum. In principle, there are numerous possibilities. However, we are are interested only in those matrices which give nonzero expectation values of the fermionic bilinears
\begin{equation}
m \; = \; \langle \Psi^\dagger M \Psi \rangle \: \neq \: 0. 
\end{equation}  
This condition can be satisfied only if 
\begin{equation}
M\;=\;- \left( \sigma_1 \otimes I_8 \right) \; M^\top \; \left( \sigma_1 \otimes I_8 \right),
\end{equation}  
where $I_8$ is the eight-dimensional unit matrix.\cite{Altland_zimbauer} It was previously shown by Altland and Zirnbauer\cite{Altland_zimbauer} that there exists a unitary matrix 
$U=U_2 \otimes I_8$, such that 
\begin{equation}
\tilde{M}=- \tilde{M}^\top,
\end{equation}
where $\tilde{M}=U M U^\dagger$. In particular one can show that \cite{herbut-cliffordalgebra}
\begin{equation}
U_2=\left( \pm i\right)^{1/2} e^{i \frac{\pi}{4} \sigma_3} e^{i \frac{\pi}{4} \sigma_2} e^{i (\frac{\pi}{4} -\phi) \sigma_3},
\end{equation}
whereas $\phi=\pi/4$ was originally considered in Ref.~\onlinecite{Altland_zimbauer}. Therefore, after the unitary rotation by $U$, the mass-matrices become purely imaginary. Transformation of the free Hamiltonian $H_0$ under the same unitary rotation can be captured by writing 
\begin{equation}
\alpha_i \;=\; Re\left( \alpha_i \right) \;+\; i \ Im\left( \alpha_i \right).
\end{equation}  
Therefore
\begin{equation}
H_k \;=\; \left( \sigma_3 \otimes Re(\alpha_i) \;+\; i \ \sigma_0 \otimes Im(\alpha_i) \right)d_i \equiv \Gamma_i d_i. 
\end{equation}
On the other hand, $U_2 \sigma_3 U^\dagger_2=\sigma_2$. After the unitary transformation the free Hamiltonian is 
\begin{equation}
\tilde{H}_k \;=\; \left( \sigma_2 \otimes Re(\alpha_i) \;+\; i \ \sigma_0 \otimes Im(\alpha_i) \right)d_i \equiv \sum^{2}_{i=1}\tilde{\Gamma}_i d_i,
\end{equation}
with $\tilde{\Gamma}_i$ now being purely imaginary matrices. A similar analysis showed that the free Dirac Hamiltonian (linear in momentum) in a monolayer graphene is defined in terms of  real $\tilde{\Gamma}_i$ matrices.\cite{herbut-cliffordalgebra}

Now we are after all the $16 \times 16$ imaginary matrices that anti commute with $\tilde{\Gamma}_i, \; i=1,2$. Since, $i \tilde{\Gamma}_j, i \tilde{M}$ are all real and square to $-1$, we first seek to know the maximal number of $q$, so that for $p \geq 0$, the dimensionality of the real representation is 16 and together close a $C(p,q)$ algebra. The answer is $8$. They form a Clifford algebra $C(0,8)$.\cite{okubo, tableigor} $C(p,q)$ defines a set of $p+q$ mutually anti-commuting matrices; $p$ of them square to $+1$, whereas $q$ of them square to $-1$. Let us now define a set of eight anticommuting real matrices  
\begin{eqnarray}
{\cal I} \;&=&\; \left( I_1, I_2, I_3, I_4, I_5, I_6, I_7, I_8 \right),
\end{eqnarray}
all of which squares to $-1$. Since, $\tilde{\Gamma}_1, \tilde{\Gamma}_2$ are imaginary, let us assume $\tilde{\Gamma}_1= i I_1$ and $\tilde{\Gamma}_2= i I_2$. Next we want to find all the Hermitian imaginary matrices, which anticommute with $I_1$ and $I_2$. The result is shown in the table below. 
\begin{center}
\begin{tabular}{r|c|c|}
\multicolumn{1}{r}{}
 &  \multicolumn{1}{c}{Mass matrix(${\cal M}$)} \hspace{0.2cm}
 & \multicolumn{1}{c}{} \\
\cline{2-3}
 & $i \; I_1 \; I_2 (\;=\; m_1)$ & $1$ \\
\cline{2-3}
 & $i  I_k  I_l  I_m  I_n I_p \; (k\neq l \neq m \neq n \neq p =3, \cdots, 8 )$ & $6$ \\
\cline{2-3}
 & $i \left( I_3, I_4, I_5, I_6, I_7, I_8  \right)$ & $6$ \\
\cline{2-3}
 & $ i m_1 \left( I_k  I_l  I_m  I_n \right)(k\neq l \neq m \neq n =3, \cdots, 8 )$ & $15$ \\
\cline{2-3}
\end{tabular}
\end{center}
The numbers in the right column indicate the number of mass matrices belongs to each class of mass matrices. Therefore, we find that there are all together $28$ imaginary Hermitian matrices (mass orders) which anti commute with the kinetic-energy Hamiltonian. This number is in accordance with the one we have computed explicitly. One can perform a similar exercise to find the number of mass matrices for single layer graphene. That number is shown to be $36$.\cite{chamon-masses}

One can also immediately find the number of gapless orders in bilayer graphene. Any gapless order, with finite expectation value, assumes the form $i \Gamma_j {\cal M}$, where $j=1$ or $2$. By construction, it is Hermitian and imaginary. Therefore the total number of gapless order parameters in bilayer graphene is $2$ (number of matrices in $H_k$) $\times 28$ (number of mass matrices) $=56$, in agreement with my explicit computation in Sec. VII.


\begin{thebibliography}{99}
\bibitem{comment-interaction} If we neglect the trigonal warping, which otherwise breaks the quadratic band touching near each valley into four Dirac cones, intrinsic screening (Thomas-Fermi screening) due to the finite density of states at Fermi energy turns the interactions into short-ranged ones. However, the vanishing density of state at the Fermi energy due to the trigonal warping, restores the long range nature of the Coulomb interaction.
\bibitem{cl3} The $Cl(3)$ group is a set of 3 mutually anti-commuting matrices. See for example Ref. \onlinecite{okubo}.
\bibitem{representation} For example one can take $\Psi \equiv \Phi$, as in Eq. (29) and (30), yielding such particular form of the kinetic energy term.
\end{thebibliography}
\end{document}